# Self-aligned pillar arrays embedding site-controlled single quantum dots for enhanced non-classical light emission

Gediminas Juska*,•, Simone Varo•, Nicola Maraviglia, John O'Hara, Salvador Medina, Luca Colavecchi, Francesco Mattana, Armando Trapala, Michael Schmidt, Agnieszka Gocalinska, and Emanuele Pelucchi

*Tyndall National Institute, University College Cork, Lee Maltings, Dyke Parade, T12R5CP Cork, Ireland*

*Corresponding author: gediminas.juska@tyndall.ie
•GJ and SV contributed equally

**Abstract**

This work presents a foundational approach for fabricating arrays of self-aligned micro- and nanopillar structures incorporating individual site-controlled quantum dots (QDs) for enhanced light extraction. This method leverages the non-planar surface morphology of pyramidal QD samples to define dielectric masks self-aligned to the QD positions. The mask size, and consequently the lateral dimensions of the pillars, is precisely controlled through a chemical mechanical polishing step, obviating the need for any additional lithography step for creating the pillar. This fabrication technique offers several key advantages, including precise control over the pillar sites, and fully deterministic embedding of QD structures. The functionality of the structures was validated by integrating single $In_{0.25}Ga_{0.75}As$ QDs – upon two-photon excitation of the biexciton state, the emission of single and polarization-entangled photon pairs was observed. Additionally, an extra fabrication step to deposit dome-like structures atop the pillars was demonstrated, effectively enhancing light extraction efficiency up to 12%.

**Keywords:** quantum dot, site-control, micropillars, entangled photons, light extraction, (111)B

**Introduction**

Epitaxially grown quantum dots (QDs) are envisioned as sources of photonic and spin qubits. However, large-scale applications require them to meet a stringent set of criteria[1]. Among these, site-control is often an overlooked feature, despite its potential to address scalability challenges. These challenges arise, in part, from the random QD nucleation sites characteristic of the most widely studied multiple self-assembled QD systems. Although site-controlled QDs face challenges such as defects introduced during site preparation and their often non-planar geometry[2], they have been successfully demonstrated as high quality emitters of indistinguishable single photons[3] and entangled photon-pairs[4]. Additionally, they have shown potential as hosts for spin qubits[5,6].

So far, photon extraction efficiency – another requisite for deterministic photon emitters – has been successfully addressed in site-controlled QD systems only by nanowires fabricated via a bottom-up method[7], each embedding a single QD. This approach has achieved a peak extraction efficiency of 30% at the first lens. In contrast, other site-controlled systems continue to face challenges in this area. Therefore, advancements in photon extraction efficiency are particularly relevant and needed.

Here, we present an innovative method for top-down fabrication of pillar-like structures embedding (111)B oriented site-controlled QDs, also known as pyramidal QDs, due to their recessed tetrahedron seeding sites. The deterministic fabrication of pillars embedding individual pyramidal QDs aims to address the extraction efficiency challenge for this class of QDs[8,9] while providing a potential route toward photonic circuit integration. Pyramidal QDs are particularly attractive due to their engineering flexibility, reproducibility and uniformity[10,11]. The fabricated micro- and nano-pillars primarily act as waveguides limiting the spatial emission modes to those propagating along the vertical (optical) axis, effectively enhancing light collection efficiency. As discussed in the following sections, a key advantage of our fabrication method is its self-aligning nature. By leveraging the intrinsic non-planarity of the system, the process eliminates the need for additional lithographic steps, thereby reducing the requirement for stringent alignment tolerances and the cost and complexity of fabrication.

The fabrication of pillar structures, such as wires and cylindrical Fabry-Pérot microcavities, has proven to be a very effective strategy to extract light, in some cases with minimal compromise regarding other important QD emission properties, such as indistinguishability and single photon purity [12, 13,14]. The majority of these systems lack the site-control capability and rely on either random site formations of (micro-/nano-)wires, pre-selection and aligning to a single QD, or just a random coincidence of optimal conditions. Also, state-of-the-art top-down methods that rely on QD pre-selection for the integration into photonic nanostructure very often face challenges related to insufficient positioning accuracy[15]. QDs misaligned from the symmetry axis can lead to polarization anisotropy[16], reduced strength of the light-matter interaction[17], and coupling to potential decoherence channels[18].

In contrast, the fabrication process shown herein ensures that the pillars host QDs precisely aligned along the central axis. Thus, this method and the future optimization of these pillar structures will represent a significant advancement in the site-controlled QD-based nanostructures for quantum information processing.

**Pillar fabrication**

The presented pillar fabrication technique exploits the non-planarity of the pyramidal QD system. The non-planarity originates from the initial template – a (111)B oriented GaAs substrate patterned with predefined arrays of tetrahedral recesses with, in general, a 7.5 to 12.5 μm pitch. The substrate patterning process, based on a crystallographically selective etching, exposes (111)A oriented surfaces, and defines a sharp tip (≤ 10 nm radius) at the bottom of each recess (see Fig. 1). These sites serve as templates for QD formation during metalorganic vapor phase epitaxy growth, enabling precise control over key structural parameters[19].

Notably, the presented pillar fabrication method is compatible with a wide range of pyramidal QD structures, including $In_xGa_{1-x}As/GaAs$[10], $GaAs/Al_xGa_{1-x}As$, GaAs/superlattice[11], and stacked-coupled QDs[20]. This versatility allows for extensive tunability in QD material composition, confinement barriers, thickness, and lateral dimensions, as well as the deterministic stacking of multiple reproducible QDs – a level of engineering control that is challenging to achieve in other QD systems. Different types of pyramidal QDs exhibit substantially different electronic structures and optical properties, each optimally suited for specific applications. In this work, we focus on $In_{0.25}Ga_{0.75}As$ QDs confined by GaAs, that have been demonstrated as highly reproducible emitters of polarization-entangled photon pairs[4,21].

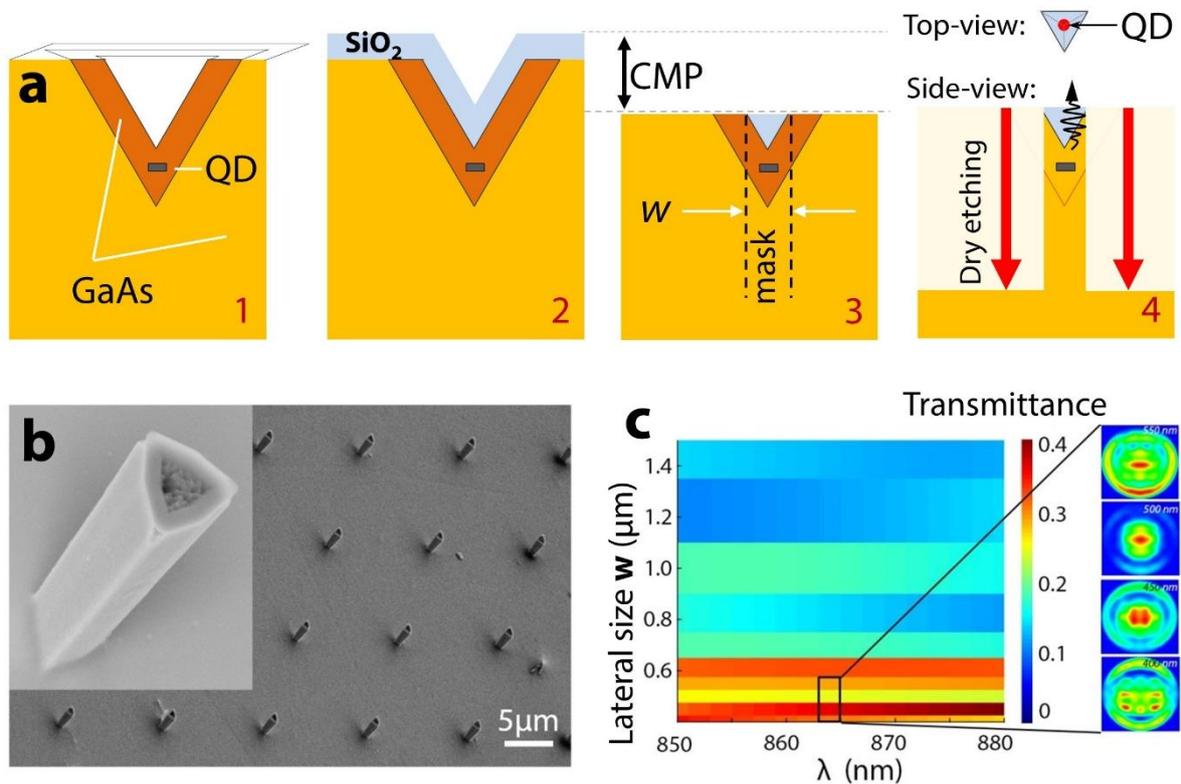

**Figure 1.** Schematics of the fabrication of the pillar structures. (a) Four key fabrication steps shown in a side-view perspective: 1) QD growth on the 111(B) oriented substrate pre-patterned with tetrahedral recesses, 2) Deposition of a $SiO_2$ layer, 3) CMP to remove the planar dielectric layer and control the triangular footprint of the remaining $SiO_2$ inside the recess, with a lateral size $w$, and 4) Formation of pillar structures from the areas protected by the remaining dielectric mask during the dry etching process. The top-view perspective shows the triangular footprint of the fabricated pillar, with a projected QD position at the circumcenter of a triangle. b) Representative SEM image of fabricated pillars. c) Numerical simulation of nanopillars. Upward transmitted power as a function of the lateral size $w$. Together with a strong dependence of the light extraction, the far field profiles change drastically. Four representative far field profiles are shown on the right of the intensity heatmap (from bottom to top, 400 nm, 450 nm, 500 nm and 550 nm width pillars). In these simulations the QD is placed 900 nm below the tip of the oxide cap.

Fig. 1a illustrates the pillar fabrication process from a cross-sectional perspective. The epitaxial growth of the sample is adjusted to retain a residual recess. Although faceting along different crystallographic directions has been observed during growth[22], the footprint of the post-growth recess ideally maintains threefold symmetry, with the rotational center aligning vertically with the underlying QD. We exploit this recess to define a self-aligned dielectric hard mask, triangularly shaped, that fills the recess and is therefore centered along the rotational symmetry axis and aligned with the QD located at the circumcenter of the triangle (see step 4 in Fig. 1a). This is achieved, firstly, by conformally depositing a $SiO_2$ dielectric layer (typically in the range of 200 – 800 nm) utilizing plasma-enhanced chemical vapor deposition (PECVD). Secondly, chemical mechanical polishing (CMP) is employed to remove the $SiO_2$ from the planar regions of the sample surface. This process also defines the footprint size of the remaining recess filled with dielectric. Polishing for a longer time results in a smaller size of the remaining recess (see step 3 in Fig. 1a). Monitoring of the remaining dielectric amount is conducted using an optical microscope or, for finer control, a scanning electron microscope (SEM) – see Fig. S1 in Supporting Information. Finally, reactive ion etching is utilized to vertically etch GaAs. At this stage, the dielectric-filled recesses act as a hard mask, controlling the lateral dimensions of the pillars. This results in an ordered array of pillar structures with triangular cross-section and a typical height of 2-4 µm, determined by the etching duration (Fig. 1b). See Supporting Information for detailed procedures.

The presented fabrication concept offers a clear path for obtaining optimised designs. Similarly to other lithographically defined pillar fabrication methods[23,24], control over critical geometric properties – such as lateral dimensions, height, and tapering – can be adopted to improve optical properties. This is supported by finite-difference time-domain (FDTD) simulations, which demonstrate a strong dependence of both the emitted light intensity and the far-field profile on the pillar width (see Fig. 1c) and the QD position within the pillar (see Supporting Information for more details and results).

Consistent with previous studies[17], our simulations show that pillars with smaller lateral dimensions exhibit higher light extraction efficiency, as such structures typically support fewer optical modes outside the vertical (optical) axis direction. For example, in cylindrical micro and nano-wires the optimal ratio between the diameter and the wavelength, where only a single fundamental mode is supported, was found to be 0.23 [17]. However, achieving this theoretical efficiency generally requires lateral dimensions smaller than those targeted in our experiments. A common challenge is that optical quality deteriorates as size decreases due to the proximity of etched surfaces' defects to the QD, leading to spectral broadening, QD charging, and non-radiative recombination.

In our pillars, the situation is even more complex than in the typical cylindrical geometry found in other systems, due to the three-fold rotational symmetry, in general, supporting a different set of modes[25], and the nonplanar termination of the top profile. As a result, FDTD simulations show that both the emission intensity and its far-field profile are highly sensitive to the pillar's dimensions and the QD position relative to the nonplanar top surface (see Fig. 1c and Supporting Information for further analysis). The dependence on the QD position is the result of the coherent interference between the multiple excited pillar modes, which is also predicted for pillars with a flat GaAs/air interface. The non-planar GaAs/SiO$_2$ interface plays a dual role: first, it introduces a different material boundary, reducing back reflection compared to a GaAs/vacuum interface; second, its specific shape generates a more complex far-field emission pattern, which can limit the amount of light collected within a restricted numerical aperture (NA).

It should also be said that our preliminary simulation results should be considered at this stage only for guidance. While they are based on realistic, experimentally characterized pillar geometries, in this phase, we were unable to unambiguously compare experimental and theoretical analysis due to the lack of a proper pillar tracking scheme during our initial developments.

**Optical properties**

The pillar structures fabricated following the above procedure showed promising results. On average, a typical light collection efficiency improvement of at least two orders of magnitude was observed, with total efficiency ranging from 0.5% to 5% at the first lens. Fig. 2a presents spectra obtained under non-resonant (635 nm) continuous-wave (CW) and pulsed (80 MHz) excitation. Experimental observations indicate a strong dependence of light collection on the numerical aperture (NA) of the objective, suggesting broad angular emission from the pillars. Notably, light collection increases several-fold when using a high-NA (0.8) objective compared to an NA of 0.42.

To probe the practical applicability and check for any lack of catastrophic processing damage, the biexciton state was initialized by two-photon excitation (TPE), which is also known to bear some challenges in this QD system. Rabi oscillations (Fig. 2b) of the biexciton (XX) and exciton (X) states were obtained, and well represent the coherent nature of the excitation process. The higher intensity of the exciton state is characteristic to these In$_{0.25}$Ga$_{0.75}$As QDs which always have an antibinding biexciton state[26]. In this case, the laser was tuned to the TPE resonance, which was energetically above the exciton state, efficiently overlapping with the asymmetric acoustic phonon distribution. Due to the interaction with the acoustic phonons, even at moderately low absolute excitation powers required to generate π pulses, the phonon-assisted exciton population ($X_{ph}$) becomes comparable to the exciton population originating from the biexciton-exciton recombination cascade as shown in Fig. 2b. The specific exciton population through the phonon-dressed state depends obviously on several parameters, such as the sample temperature, the exciton or QD dimensions, pulse temporal width and shape[27], and optimal pumping conditions limiting this effect might change from sample to sample.

The unwanted photons generated through the phonon-assisted excitation pathway can be effectively filtered out during second-order correlation function measurements relating biexciton and exciton. In these measurements, the correlation events occurring around the first excitation pulse (±6.25 ns) are exclusively linked to the biexciton-exciton recombination cascade. The purity of the single-photon emission, as confirmed by the second order auto-correlation functions for the exciton and biexciton (0.046±0.037 and 0.032±0.025, respectively), indicates good single-photon emission triggered by the excitation pulse (Fig. 2c). These photon pairs, originating from the recombination cascade, are well known to serve as a resource for polarization entanglement. However, one of the major challenges in using these photon pairs for practical applications is the fine-structure splitting ($S$) of the exciton level. This splitting typically causes the precession of the non-degenerate exciton spin state after the biexciton recombination, leading to a randomization of the overall two-photon polarization states[4,28,29].

The fine-structure splitting issue in the pyramidal QD system is known to be largely mitigated by the intrinsically high rotational symmetry[30]. Although other factors, such as QD or confinement barrier alloy disorder, may introduce deviations from this ideal model, several highly symmetric pyramidal QD designs and fabrication strategies were developed[4,11]. The thin (0.9 nm) $In_{0.25}Ga_{0.75}As$ QDs presented here, with ground bound states near the confinement barrier energy, generally exhibit an average $S$ of a few µeV (1.6±0.8 µeV for this sample). The representative QD shown in Fig. 2 has $S$= 1.3±0.4 µeV extracted from the two-photon polarization state phase oscillation observed in diagonal and circular polarization bases. By measuring the full set of linear, diagonal, and circular polarization bases for the two-photon (biexciton-exciton) intensity correlations (Fig. 2d), we calculated the fidelity to the expected maximally entangled state $\sqrt{0.5}(|HH\rangle + |VV\rangle)$ to be 0.59±0.01. This value significantly exceeds the 0.5 limit for a statistical mixture of $|HH\rangle$ and $|VV\rangle$ (which corresponds to classical light), confirming polarization entanglement.

Several factors contribute to the degradation of entanglement. First, the two-photon $|HH\rangle$ and $|VV\rangle$ intensities deviate from the ideal source, resembling the state $\sqrt{0.59}|HH\rangle + \sqrt{0.41}|VV\rangle$ due to a partially induced polarization, observed as linear polarization anisotropy from the top view (see Supporting Information for further analysis). The degree of linear polarization ($D$) was measured to be 0.18, where $D = (I_H - I_V)/(I_H + I_V)$, with $I_H$ and $I_V$ representing the biexciton transition intensities of the horizontally and vertically polarized components, respectively. Although it might be tempting to attribute the origin of this anisotropy solely to structural asymmetries in the fabricated pillars, such a conclusion cannot be made definitively. The origin of the anisotropy is likely more complex and could result from multiple contributing factors. While $D$ does increase statistically after fabrication (samples were measured before and after processing), it was already observed as non-zero in as-grown QDs, possibly also due to some heavy-light hole mixing or other yet unclear effects.

Further degradation of entanglement is obviously also related to the presence of the non-negligible fine-structure splitting, exciton spin-scattering, and cross-dephasing events.

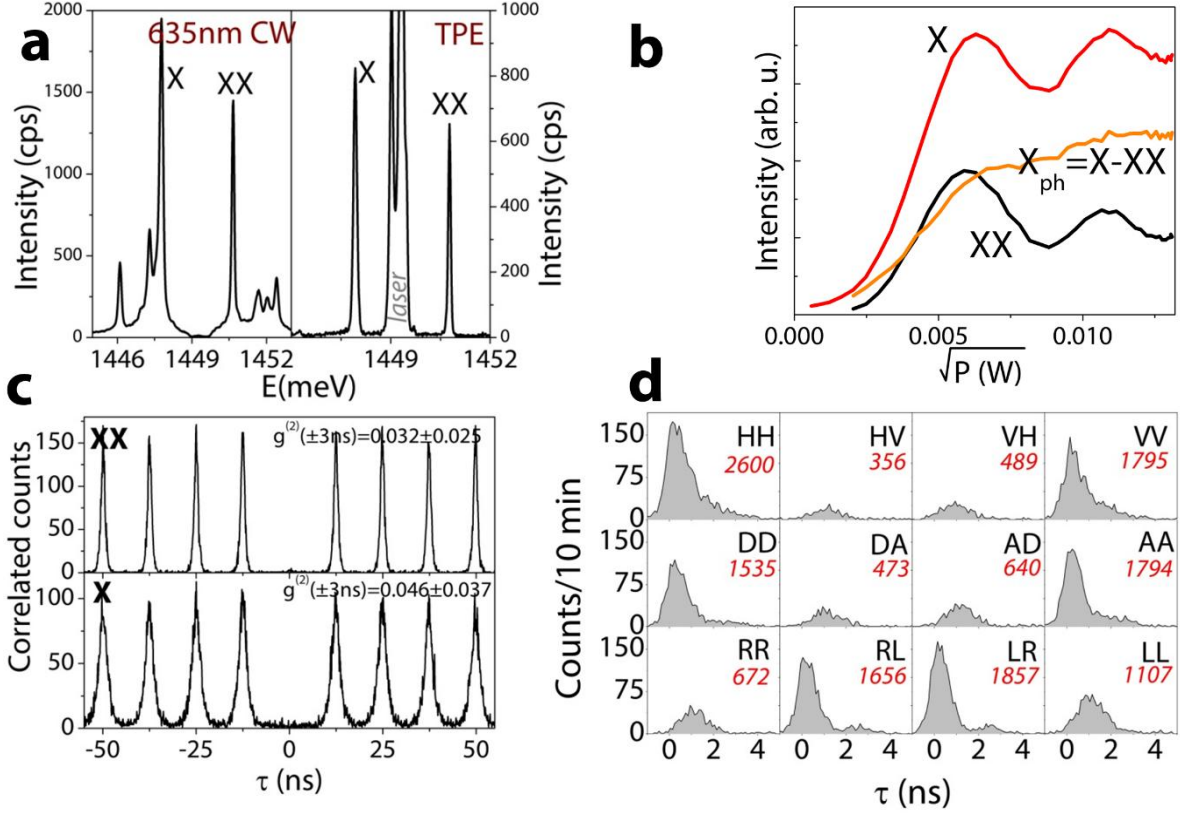

**Figure 2.** Non-classical light emission. (a) Comparison of the QD spectra under the non-resonant CW excitation and resonant two-photon excitation of the biexciton state. (b) The coherent nature of the biexciton state population is attested by Rabi oscillations as a function of the excitation power (laser pulse area). $X_{ph}$ represents the difference between the total measured intensity of exciton and biexciton transitions, corresponding to the exciton intensity component from phonon-assisted excitation. (c) Pure single photon emission of the biexciton and exciton transitions. (d) Polarization-resolved cross-correlation curves measured in linear, diagonal and circular bases sufficient to measure the fidelity to the expected maximally entangled Bell's state. The integrated counts (the two-photon projection intensity) during the 10 minutes integration time are given below the projection labels. The exciton spin state precession due to $S$=1.3 µeV is observed in diagonal and circular bases.

**Brightness optimization.**

A potential route for brightness optimization was demonstrated by introducing additional dielectric layers covering the fabricated pillars. For example, three layers of $SiN_X/SiO_2/SiN_X$ (with approximate thickness of 50 nm, 500 nm, and 300 nm, respectively) were deposited by plasma enhanced chemical vapour deposition (PECVD) (see Fig. 3b). The initial motivation for adding the dielectric layer was twofold: first, to mimic a controlled increment in the lateral dimensions of the pillar therefore modifying the optical modes supported by the pillar, and second, to create a rounded (convex) structure on the top of the pillars to induce a lensing effect for the fraction of light emitted upward. Simulations indicate that the dielectric layers alter the optimal emission geometry of the pillar, specifically in terms of the distance between the QD and the top non-planar surface. Simultaneously, the simulations show that a rounded top, modelled as depicted in the inset of Fig. 3b, provides more collimated far-field emission compared to a regular pillar, particularly when the QD is positioned at the optimal distance from the top (see Supporting Information Fig. S5).

After the deposition of the first two dielectric layers, a significant red-shift of the average emission energy of the QDs by approximately 15 meV was observed at cryogenic temperatures. This shift was likely due to the strain induced by the different thermal expansion coefficients of the materials encapsulating the pillars with respect to the GaAs one. To mitigate this effect, we chose to add a third, thick layer of $SiN_X$. Although the thermal expansion coefficients of all three materials are nonmonotonic and strongly temperature-dependent[31,32,33], qualitative estimates suggest that structural changes during the cryogenic cooling cycle are more significantly influenced by $SiN_X$ than by $SiO_2$ and are somewhat closer to those of GaAs. This $SiN_X$ addition resulted in a blue-shift in the average emission energy, partially restoring the QDs original emission energy. The dielectric coating significantly enhanced light extraction across the entire sample. Fig. 3c shows a representative spectrum of a QD obtained by scanning the transmission wavelength of a monochromator and detecting the photoluminescence signal with an avalanche photodiode (APD) at the single-photon level. The measured single-photon emission intensities of a negatively charged trion ($X^-$, 450 kcps) and a neutral exciton (X, 50 kcps) correspond to a light extraction efficiency of ~9.5% after passing through the collection objective, which has a numerical aperture (NA) of 0.42. Multiple QDs with similar light extraction values were found, with a peak value of 12%. A detailed summary and analysis of the optical losses caused by the setup elements is provided in the Supporting Information. Further investigation is needed to determine the individual contributions of the modification of the guided modes of the pillar and the lensing effect to the brightness enhancement observed in the coated pillars.

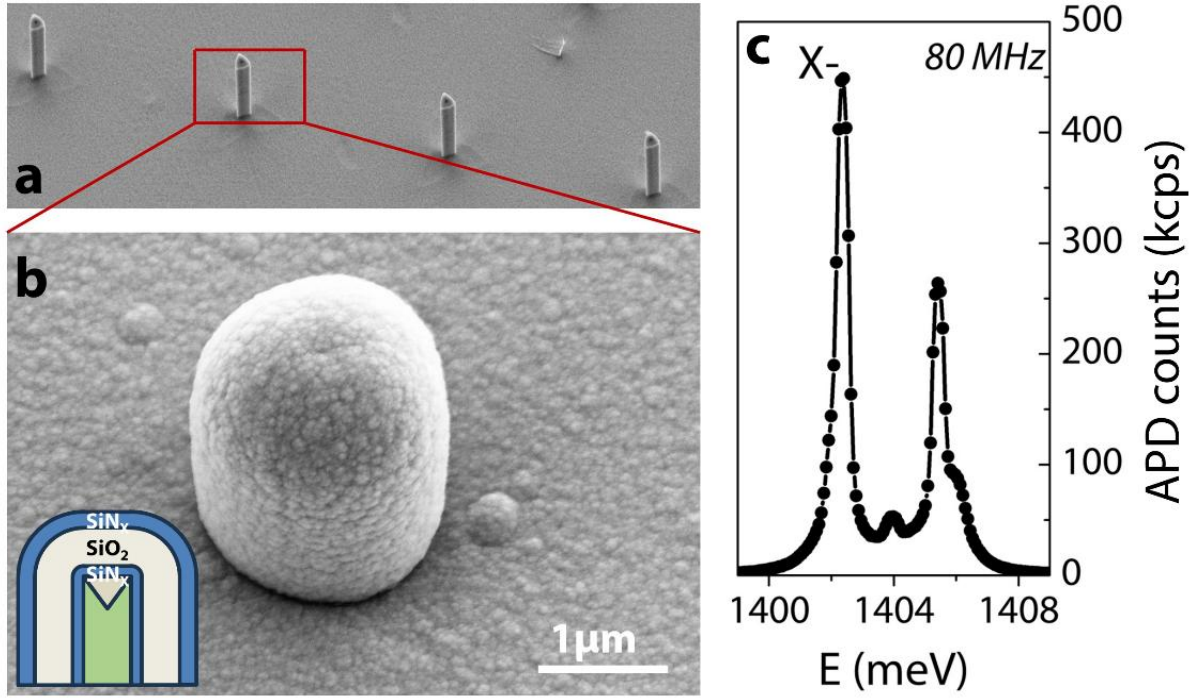

**Figure 3.** (a) The SEM image of pillars before the coating by multiple dielectric layers, (b) and the formation of dome-like structures after the deposition. (c) The photoluminescence spectrum of a QD from the fabricated structures. The spectrum was obtained with a single photon detector and indicates light extraction efficiency of at least 9.5% in this representative case.

**Discussion and Outlook**

This work lays the foundation for an innovative approach to fabricate site-controlled micro- and nanopillar structures embedding highly uniform site-controlled QDs. The key benefits of this method are: (1) deterministic control of the positions of pillars, (2) all pillars consistently containing highly accurately centred QD(s) (and potentially multiple stacked ones even if we discussed only the single QD case), and (3) achieving all of this without relying on additional lithography, registration and alignment marker steps. Moreover, both $In_xGa_{1-x}As$ QDs confined by GaAs barriers and (In)GaAs QDs confined by $Al_xGa_{1-x}As$ barriers can be embedded to exploit the main advantages of the pyramidal QD system. These advantages include high rotational symmetry for efficient polarization-entangled photon emission and engineering flexibility to design highly reproducible single and coupled QD structures. The results presented here from $In_{0.25}Ga_{0.75}As$ QDs are first demonstrators of these capabilities and their potential.

Nevertheless, to advance towards practical applications, several optimizations and developments are necessary. The demonstrated prototype, which utilizes a dielectric coating atop the pillar to enhance the lensing effect, represents one potential optimization element part of a more overall complex picture. Another key improvement would be reducing the lateral dimensions of the pillars to limit light coupling primarily to the fundamental mode.

However, this approach presents challenges due to the increased proximity of surface defect states induced by processing, making the development of effective passivation techniques critical. Several approaches are available to address this issue[34]. Additionally, the present strategy can only collect photons emitted through the pillar in the upward direction. A back reflective element would need to be engineered to increase the collection efficiency above 50%. Furthermore, the reflector could, in principle, create a micro- or nano-resonator with resonant optical modes that exhibit Purcell enhancement[17].

Even without a designed back reflector, simulations indicate the presence of an optimal QD position along the vertical axis associated with higher emission intensity (see Fig. S2b). This increase in intensity is related to constructive interference between the different modes supported by the pillar, and how these are transmitted across the top interface. Incidentally, a fraction of the light is also reflected back into the pillar and simulations show a moderate Purcell enhancement. The latter can be further boosted by optimising the interface for back reflection.

Due to the non-planar top surface of the pillar, we expect far-field distortions that depend on the QD position along the axis, as shown in Fig. S2b (the effect for different NA values is presented in Supporting Information Fig. S5). This distortion could potentially be mitigated by developing a planarization process, possibly in combination with modified etching techniques to fabricate tapered structures (Fig. S6). Tapered pillars could potentially be detached and flipped to ensure an adiabatic mode transition to a near-Gaussian beam profile[35], improving both light extraction and directionality.

Once optimized, a highly attractive prospect is the physical manipulation of these pillars. In Fig. S7, we demonstrate a preliminary procedure for transferring a pillar onto a pre-selected surface. This opens up possibilities for positioning the pillars in alignment with waveguides, fiber cores, and similar hybrid photonic platforms.

**Associated content**

**Supporting Information.** Additional information on processing details, optical set-up characterization, polarization anisotropy analysis, simulation details, and multiple engineering perspectives.

**Acknowledgement**

This work is supported by Research Ireland, previously known as Science Foundation Ireland under grants, 15/IA/2864, 22/FFP-A/10930, 18/SIRG/5526, 22/FFP-P/11530, 12/RC/2276 P2, and Enterprise Ireland under grant DT 2019 0090B.

# Supporting Information

**Processing details**

The fabrication of micro/nano-pillars begins with the deposition of a conformal $SiO_2$ layer, ≤1 μm thick, using plasma-enhanced chemical vapor deposition (PECVD). This layer covers both the recesses and the (111)B flat surface between the QDs. To expose the flat GaAs surface, an extended polishing step is performed using a Logitech PM5 grinder/polisher. The sample is bonded to a circular quartz glass slide using wax, positioned face-down at the bottom of a steel jig, and held in place by vacuum. The surface is then lapped against a soft pad attached to a polishing plate, while polishing slurry (Pureon - Ultra SolEX), containing silica particles ~80 nm in diameter, is periodically applied to the pad. During the CMP, the $SiO_2$ is removed leaving it only in the recesses until the targeted lateral dimensions (~1 μm or less) are reached (Fig. S1). After polishing, the sample is rinsed with acetone, followed by IPA, and dried with nitrogen to remove any residual organic material.

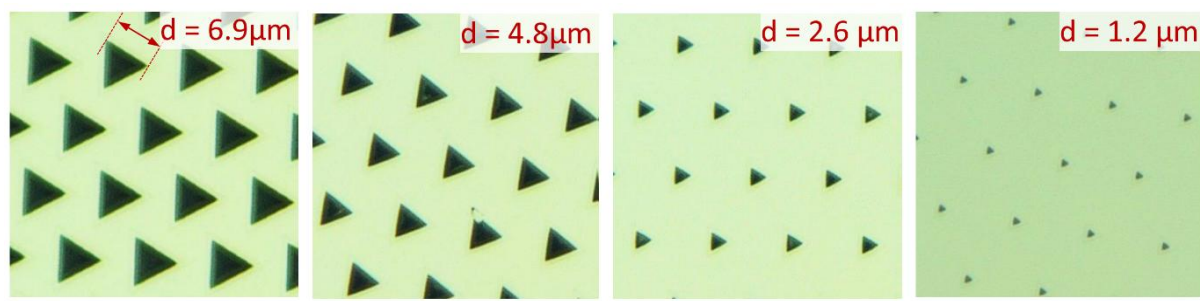

**Figure S1.** Optical microscopy images showing the top view of samples after varying polishing durations, resulting in a controlled reduction of the $SiO_2$-filled recess dimensions. The side length of the recess is denoted by *d*.

Once cleaned, the sample undergoes a dry etching process. The employed chemistry and ion energies are highly selective towards GaAs, allowing for the nearly vertical etching of GaAs while leaving the $SiO_2$ hard mask mostly unaffected. Plasma is generated from a mixture of $BCl_3$, $Cl_2$, Ar, and $N_2$ (13, 13, 13, and 5 sccm, respectively)[36], under a chamber pressure of 5 mTorr, with an Inductively Coupled Plasma (ICP) power of 600 W and an RF power of 55 W. The addition of Ar helps to stabilize the plasma and improve surface roughness by sputtering non-volatile products[37,38], while $N_2$ enhances the dissociation of $BCl_3$ and aids in the passivation of sidewalls to reduce lateral etching. The dry etching process is carried out using a PlasmaPro 100 Cobra ICP system.

**Optical setup**

The photoluminescence measurements were performed in a standard free-space micro-photoluminescence setup at 13 K. Non-resonant excitation was performed by a 635 nm temperature stabilized laser diode operating at CW or 80 MHz. The resonant two-photon excitation was performed by 10 ps energy tunable pulses obtained from a 4f pulse shaping system with Tsunami Ultrafast Ti:Sapphire oscillator at 80MHz.

Table 1 lists all relevant elements contributing to the optical losses. Due to the accumulated losses, the overall setup efficiency during the light collection measurements was 5.3%. The photon extraction efficiency measurement was based on a direct single photon detection count by a silicon Avalanche Photodiode (APD) with dead-time of 80 ns. The detector counts were registered by the Time Tagger Ultra TCSPC module from Swabian Instruments.

**Table 1. Efficiency of the optical elements at 850 nm**

| Set-up element | Efficiency |
| --- | --- |
| Cryostat window | 0.95 |
| Objective | 0.7 |
| Beam splitter (90:10) | 0.86 |
| 6x Mirror (protected Ag) | 0.89 |
| Long-pass filter (FEL700) | 0.98 |
| 2x Lens (B coating) | 0.90 |
| Grating 950/mm | 0.33 |
| Coupler + MM fiber | 0.80 |
| Detector (APD) | 0.45 |

**Simulation details**

Simulations of the transmission values and far-field profiles were conducted using the Ansys Lumerical FDTD software. The simulated structures were modelled as a pillar with perfectly vertical walls, terminating with an ideal tetrahedron recess, filled by a 200nm thick layer of $SiO_2$ on the sidewalls of the recess. The final dimensions of the $SiO_2$ cap are adjusted accordingly with the dimension of the pillar width resulting in a small empty recess left on top of pillars with larger dimensions (when the lateral size is more than 730nm). Fig. S2a presents a cross-sectional map of the refractive indices for the simulated GaAs pillar structure at a cryogenic temperature of 12 K. The QD emission is modelled as a linearly polarised dipole source.

In the context of the simulation software, the power transmission from the pillar at each wavelength is defined as the ratio of the power collected by a monitor placed 20nm above the pillar top to the power the source would emit in a homogeneous medium. Under this definition, the transmitted power can exceed a value of one, which arises from Purcell enhancement of the source. This enhancement (or suppression), even in the absence of reflections from the bottom surface, can occur due to constructive interference conditions at the dipole position, caused by reflections from the top surface[39].

The Purcell factor is defined as the ratio of the power emitted by the source in the designed medium to the power emitted in a homogeneous medium (bulk material). It is important to note that the Purcell factor is not automatically used as a normalization factor for the transmission values.

In each simulated scenario, a monitor positioned 10 nm above the simulated structure collects the emitted power within a numerical aperture (calculated from the bottom of the tetrahedron recess) NA > 0.98. The body of the pillar intersects an absorption boundary of the simulation region 1 µm below the dipole position. This configuration balances simulation accuracy and computational efficiency, allowing for reasonably long simulations while enabling precise far-field decompositions. The wide collection angle, resulting from the monitor proximity, further enhances the accuracy of these calculations.

Far-field projections are computed by the simulation software by performing a Fourier decomposition of the field data collected at the monitor into a set of plane waves propagating at different angles. The far field is then processed with a MATLAB script to determine the fraction of light transmitted at different angles with respect to the optical axis of the system. This analysis helps to evaluate the extent to which the QD emission approximates a Gaussian profile, a desirable property for integrated photonics applications.

Near-field effects are not considered in this analysis, as the decomposition into plane waves is limited to the wavelengths specified for the source and excludes rapidly varying spatial components.

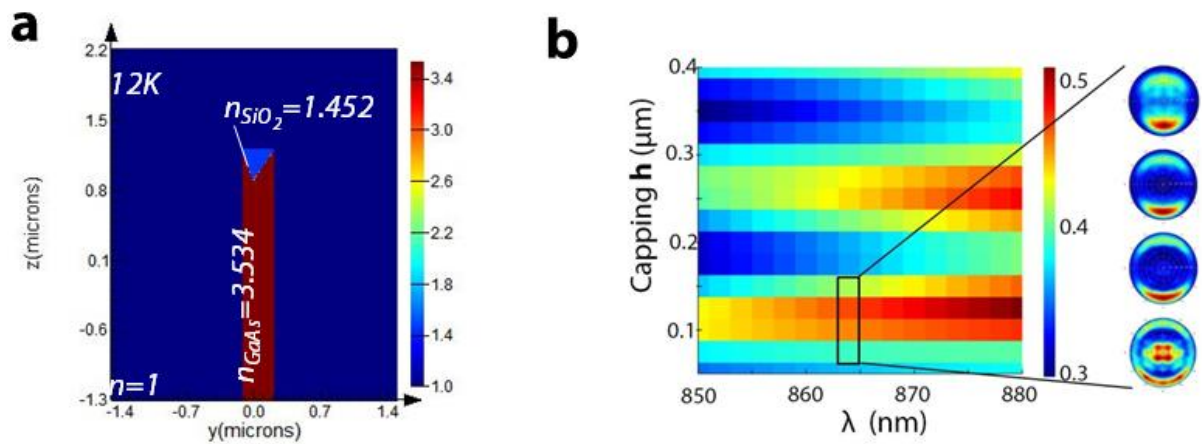

**Figure S2.** (a) Refractive indices map of the simulated pillar structure. The values are set for the experimental conditions at T=12 K. (b) Heatmap of transmission as a function of the thickness of the GaAs capping layer (h) on top of QD, determining the final distance between QD and $SiO_2$ filled recess. Four far-field profiles are shown on the right. Pillar width is 450nm.

Fig. S2b illustrates the transmission as a function of the QD capping thickness. Regions of increased transmission correlate with Purcell enhancement caused by reflections from the top surface. The far-field profiles within these high-transmission regions exhibit a broad angular distribution. Similar to the results obtained from simulations of the transmission dependence on pillar thickness, these findings demonstrate that the optical properties of the fabricated pillars are highly sensitive to both the epitaxial design and the resulting geometry of the pillars.

**Polarization anisotropy**

Fig. S3 shows linear polarization analysis of the exciton (X) and biexciton (XX) transitions from multiple pillars. Non-zero degree of linear polarization (*D*) was common to all measured structures (Fig. S3b). At this stage of the research, it was impossible to identify whether the polarization anisotropy was inherited from the as-grown QD, or was introduced during the processing. Due to lack of a systematic QD tracking/mapping system, we could not measure precisely the *D* change after the dry etching procedure, though an overall higher statistically observed *D* value suggests the contribution from the fabrication process.

To understand if the polarization anisotropy is related to the threefold rotational symmetry of pillars, the distribution of linear polarization axes angles was measured (Fig. S3a shows H linearly polarized component orientation angle). Lack of clear bunching around the values associated with the triangular geometry (for example, QD elongation/asymmetry along one of the three equilateral axis) suggests another reason. These could be subtle geometry asymmetries of the pillar, however, photoluminescence data from individual pillars must be related to their structural analysis by electron microscopy.

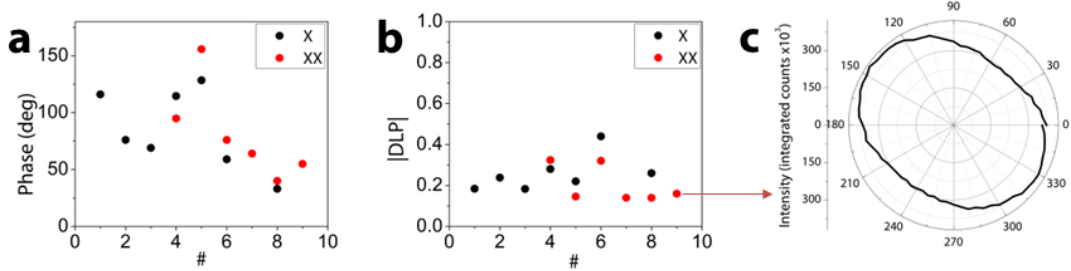

**Figure S3.** (a) Polarization anisotropy axis angle taken at the maximum of $I_H$. (b) Degree of linear polarization for the exciton and biexciton transitions. (c) A representative QD with $D_{XX}=0.18$.

**Effects of linear polarization anisotropy**

To understand how linear polarization anisotropy affects entanglement, we calculated the dependence of concurrence, as an entanglement witness, on the degree of linear polarization. We considered a two-photon polarization state:

$$|\psi\rangle = a|HH\rangle + b|VV\rangle, \qquad (1)$$

where *a* and *b* are probability amplitudes satisfying the normalization condition. The probability amplitudes can be expressed by the degree of linear polarization *D*:

$$a = \sqrt{(1+D)/2}, \qquad (2)$$

$$b = \sqrt{(1-D)/2}, \tag{3}$$

where $a^2 = I_{HH}$ and $b^2 = I_{VV}$. In the experimental measurements shown in Fig. 2, the two-photon intensity ratio was measured to be $I_{HH}/I_{VV} = 2600/1795 = 1.45$ corresponding to $D = 0.18$. Similarly, the single-photon intensity measurement obtained from the biexciton transition, initialized by two-photon excitation, yields $I_H/I_V = 1.37$, corresponding to $D = 0.16$ suggesting that $I_{HH} \approx I_H$ and $I_{VV} \approx I_V$ possibly due to the correlation between biexciton and exciton polarization states.

The two-photon density matrix is given by

$$\rho = |\psi\rangle\langle\psi| = \begin{pmatrix} a^2 & 0 & 0 & ab \\ 0 & 0 & 0 & 0 \\ 0 & 0 & 0 & 0 \\ ab & 0 & 0 & b^2 \end{pmatrix}. \tag{4}$$

It is convenient to quantify entanglement by the concurrence[40] $C$. It is defined as $C = max\{0, \lambda_1 - \lambda_2 - \lambda_3 - \lambda_4\}$, where $\lambda_i$ are the square roots of eigenvalues of the non-Hermitian matrix $\rho\tilde{\rho}$, ordered decreasingly. $\tilde{\rho}$ is the spin-flipped state expressed as $\tilde{\rho} = (\sigma_y \otimes \sigma_y)\rho^*(\sigma_y \otimes \sigma_y)$, where $\sigma_y$ and $\rho^*$ are the Pauli operator and the complex conjugate of the matrix $\rho$, respectively. For a pure two-qubit state (1), the concurrence can be calculated as $C = 2ab = \sqrt{1-D^2}$. Fig. S4 illustrates this dependence. For a specific studied and a typical case of $D = 0.18$, despite significant polarization anisotropy ($I_{HH}/I_{VV} = 1.45$), concurrence maintains a high value of 0.984.

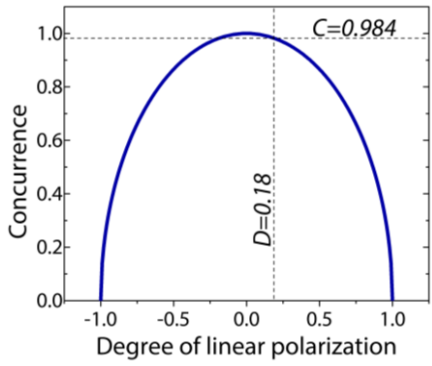

**Figure S4.** Concurrence (C) as a function of the degree of linear polarization (D). The specific measured case with D=0.18 is highlighted on the graph.

### Engineering perspectives: dielectric coating

Fig. S5 illustrates the effect of pillar coating with a $SiN_X/SiO_2/SiN_X$ stack, with thicknesses of 50 nm, 480 nm, and 310 nm, respectively (Fig. S5a). The GaAs pillar width is set to be 500nm. To investigate the impact on transmitted power, the QD vertical position $h$ relative to the bottom of the recess was varied, and the resulting transmission was analysed for numerical apertures (NA) of 0.2, 0.45, and 0.7.

The results clearly demonstrate an overall increase in collected transmission power across the studied NA values. Additionally, far-field profile analysis (Fig. S5d) reveals a more directional emission pattern, which correlates with the observed increase in intensity from the coated pillar structures.

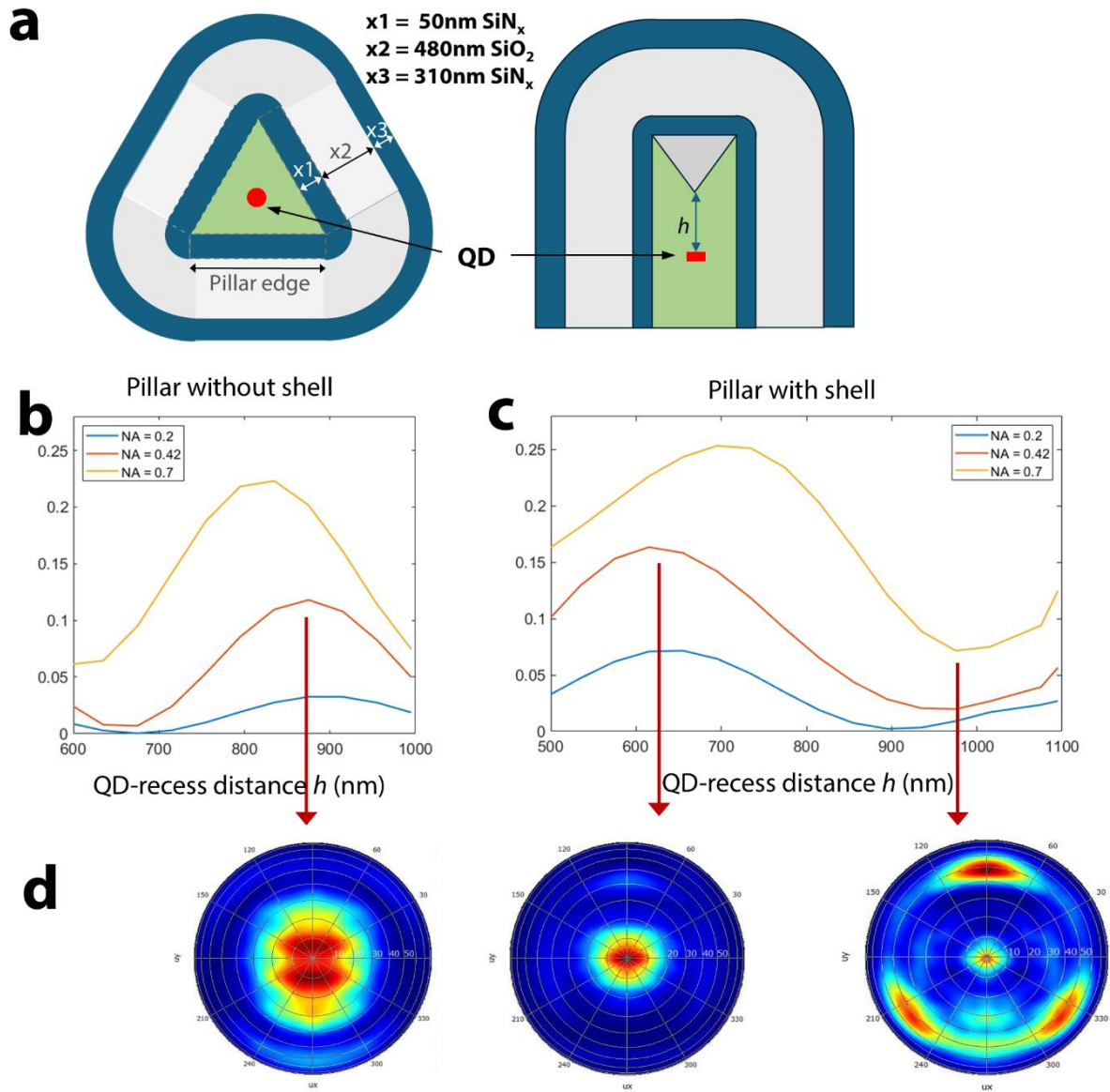

**Figure S5.** (a) The simulated structure geometry with indicated thickness of dielectric layers. Comparison of power transmission from pillars (b) without and (c) with dielectric coating as a function of the QD distance from the recess. Transmission curves are presented for three numerical aperture (NA) values: 0.2, 0.42, and 0.7. (d) The corresponding far-field profiles at the indicated points are shown below.

**Engineering perspectives: pillar tapering**

Fig. S6 illustrates the possibility of achieving pillar tapering through a modified dry etching process.

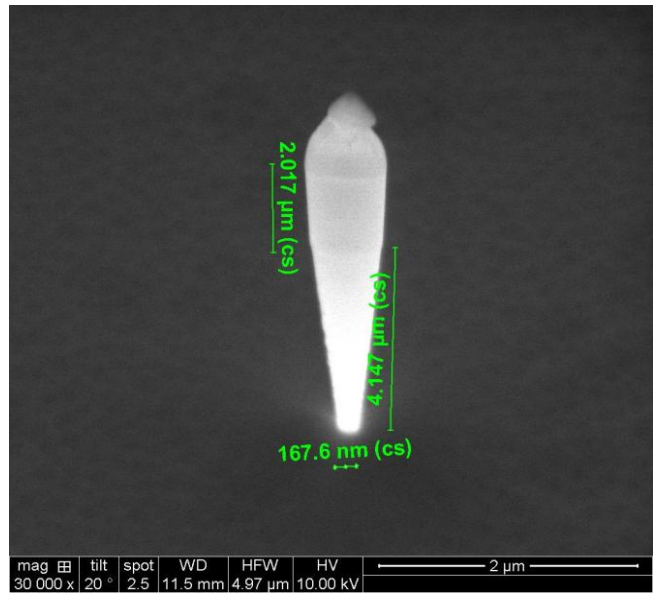

**Figure S6.** A tapered pillar.

**Engineering perspectives: pillar transfer**

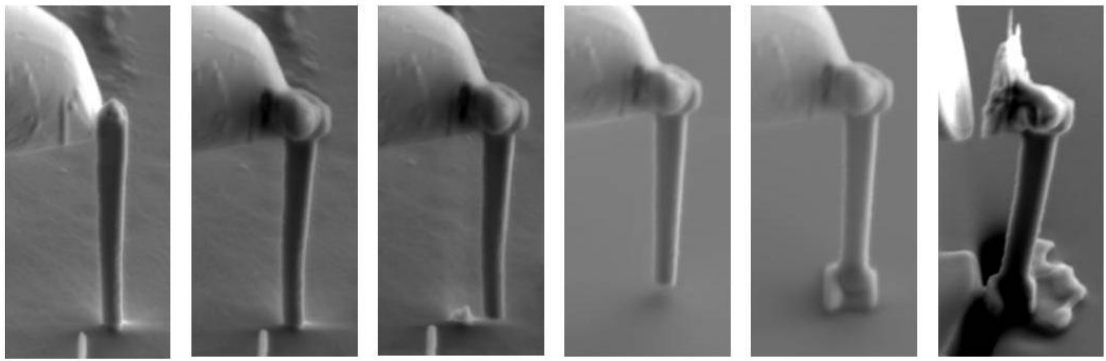

**Figure S7.** A pillar transfer procedure.

To probe further processing versatility, the nanopillar was transferred in a Tescan Solaris FIB-SEM. It was attached to an Omniprobe needle using e-beam deposited $SiO_2$. No FIB deposition was used to avoid Ga ion implantation. The nanopillar was then sheared off at the base by pushing the Omniprobe needle against the pillar. Then the base of the pillar was secured to the optical fibre by another e-beam deposited $SiO_2$. Finally, the tip of the Omniprobe needle was cut off using the ion beam, so the nanopillar is standing free again.

---

[40] Wootters, W. K., Entanglement of Formation of an Arbitrary State of Two Qubits, Phys. Rev. Lett. 80, 2245, 1998.